\documentclass[final,3p]{elsarticle}

\usepackage{amssymb}

\usepackage{color}

\journal{C. R. Physique}

\begin{document}

\begin{frontmatter}



\title{Rheological properties of dense granular flows}


\author{Pierre Jop}

\address{Surface du Verre et Interfaces, UMR125 CNRS/Saint-Gobain,

33 quai Lucien Lefranc, 93303 Aubervilliers}
\ead{pierre.jop@saint-gobain.com}

\begin{abstract}

\noindent Recent progresses in understanding the behavior of dense granular flows are presented. After presenting a bulk rheology of granular materials, I focus on the new developments to account for non-local effects, and on ongoing research concerning the surface rheology and the evolution of mechanical properties for heterogeneous systems.

\vskip 0.5\baselineskip
\noindent{\bf R\'esum\'e}
\vskip 0.5\baselineskip
\noindent
\textbf{Propri\'et\'ees rh\'eologiques des \'ecoulements granulaires denses.} Les r\'ecents pro-gr\`es de compr\'ehension des \'ecoulements granulaires denses sont pass\'es en revue en se focalisant sur la rh\'eologie classique des mat\'eriaux granulaires tous en raffinant notamment sur les effets non-locaux, l'influence des conditions limites des \'ecoulements et l'\'evolution des propri\'et\'es d'\'ecoulements de syst\`emes h\'et\'erog\`enes. 
\end{abstract}


\end{frontmatter}
\pdfinfo{
   /Author (Pierre Jop)
   /Title  (Rheological properties of dense granular flows)
   /CreationDate (D:20141220145631)
   /ModDate (D:20141220145631)
   /Subject (Review on the rheology of granular flows)
   /Keywords (dense granular flow;rheology;non local effects;boundary condition;review)
}
\section{Introduction}


Granular materials have been extensively studied in particular due to their large involvement in geophysical events and in industrial processes.
\textcolor{black}{From powders to boulders avalanches, the granular systems are composed of individual grains interacting through solid contacts and they share some features}. Compared to classical Brownian particles or molecules, their relatively large size ($d>100$ $\mu$m) prevents thermal energy from significantly sustaining  agitation or leading to an equilibrium state. Thus external mechanical energy must be supplied to shear or to deform the system.

Depending on the amount of energy, three qualitatively different states can be distinguished \cite{Jaeger1996}.
When the grains have very little kinetic energy, the system is dense, the structure is dominated by force chains and the behavior is well described by the soil mechanics. On the other hand, if a lot of energy is brought to the grains, the system is dilute, grains interact only by binary collisions and the flow properties share common features with an agitated gas.

\begin{figure}
\centering
\includegraphics[scale=1.5]{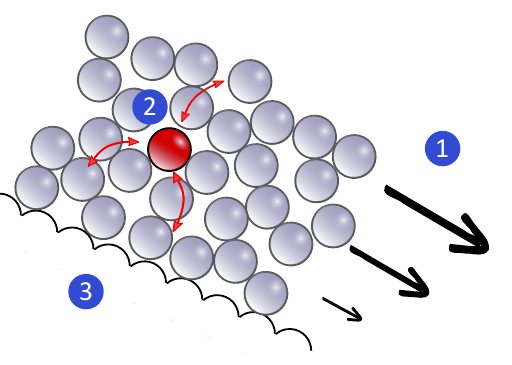}
	\caption{Schematic of a flow on inclined plane illustrating three issues to describe the dense granular flows. First, what can predict a \textcolor{black}{frictional rheology} for the bulk flow? Second, when are the coupling across space important? Third, which interaction law governs the flow properties at boundaries?  
	}
	\label{fig:denseflow}
\end{figure}

However, due to the dissipative nature of these interactions, the loss of energy may decrease the kinetic energy leading rapidly to a denser system as often encountered in common granular flows. In such flows, grains have enduring long-lasting contacts with their neighbors and mostly interact through frictional and collisional forces. Despite the large interest for granular flows and the recent advances in modeling, a general theoretical framework for a granular rheology covering all the situations is far from been achieved.
In this article, I will mainly focus on the rheological properties of spherical hard grains interacting through dry frictional contacts. One example is the flow on inclined plane as illustrated in figure (\ref{fig:denseflow}) where I split the problem of describing the flow into three numbered issues. The first one concerns the \textcolor{black}{lowest-order} laws able to model the velocity profile in the bulk. The second one deals with spatial correlations and their consequences on the rheology. The third one aims to describe the interactions between the flow and the boundaries. These issues will guide the discussion along this article.

In the following, I will first present the bulk behavior of the granular matter and a rheological law (section \ref{sec:local}).
Since deviations from its predictions exist, I will then show different ways to include "non-local" effects to extend the previous theoretical description (section \ref{sec:nonlocal}).
In section \ref{sec:surface}, I will make an overview of the surface properties issues that arise in real systems and the different models trying to incorporate them.
In the last part (section \ref{sec:discussion}), I will discuss the evolution of this model when dealing with other geometries of grains or interactions.

\section{Bulk rheology}
\label{sec:local}
\textcolor{black}{In this section, I will assume that the local friction coefficient does not depend on higher order derivatives of the velocity field than the local stain rate. For sake of simplicity, I will call this rheology a \textit{local} rheology. On the contrary, if higher-order derivatives in space are involved or if other fields modifies the lowest-order behavior leading to a stronger coupling over distance, we cannot define a unique flow curve linking the shear stress to the strain rate (and to the pressure). I will call it a \textit{non-local} rheology in this article and I will discuss this point in the section \ref{sec:nonlocal}.}

\subsection{Properties of steady dense flows}
I briefly describe here the basic features of steady dense granular flows in simple geometries. To have a complete overview of flow properties both in experiments and in numerical simulations, readers may refer to some recent reviews \cite{
MiDi2004,Delannay2007,Forterre2008}.

From a rheological point of view, the simplest configuration is the plane shear geometry. The granular layer is only confined by an external \textcolor{black}{constant} pressure $P$ and sheared by driving one boundary either at constant velocity or at constant shear stress $\tau$. In steady state, simulations have shown that the velocity profile is almost linear except close to the boundaries \cite{Iordanoff2004,Dacruz2005}.

Experimentally it is more convenient to study surface flows. %
The inclined plane configuration has the advantage to produce uniform flows, \textcolor{black}{for suitable parameters},  with only two control parameters: the inclination of the plane ($\theta$) and the thickness of the flowing granular layer $h$. Above a thickness threshold ($h_{\textrm{\scriptsize{stop}}}(\theta)$) the material flows and exhibits steady uniform flows in a \textcolor{black}{narrow range of inclinations of few degrees (5$^o$ or 10$^o$) \cite{Pouliquen1999,Forterre2003,Borzsonyi2009}}. For high enough thicknesses, the mean velocity scales with the thickness to the power $3/2$ and is inversely proportional to the thickness of the deposit $h_{\textrm{\scriptsize stop}}(\theta)$. This scaling is known to fail close to the threshold, for thin layers \cite{Rajchenbach2003,Silbert2003}, I will discuss this point in section \ref{sec:nonlocal}. These observations were identical in numerical simulations \cite{MiDi2004}.%
 When the grains are not flowing on a plane but on top of a pile, the convexity of the velocity profile changes presenting a fast flowing layer and a creeping flow deep in the pile \cite{Komatsu2001}. In this configuration, the slope is selected by the system and the flow rate \cite{Taberlet2003}. 
 Finally, in confined configurations such as the Couette cylinder or the silo where the stress spatial distribution is highly heterogeneous, the flow exhibits shear localization close to their boundaries \cite{MiDi2004}.

Some of these behaviors can be gathered in the same theoretical framework. In the following section, I will introduce the main ideas of this rheological model.

\subsection{Dimensional analysis}
\begin{figure}
\centering
\includegraphics[scale=.3]{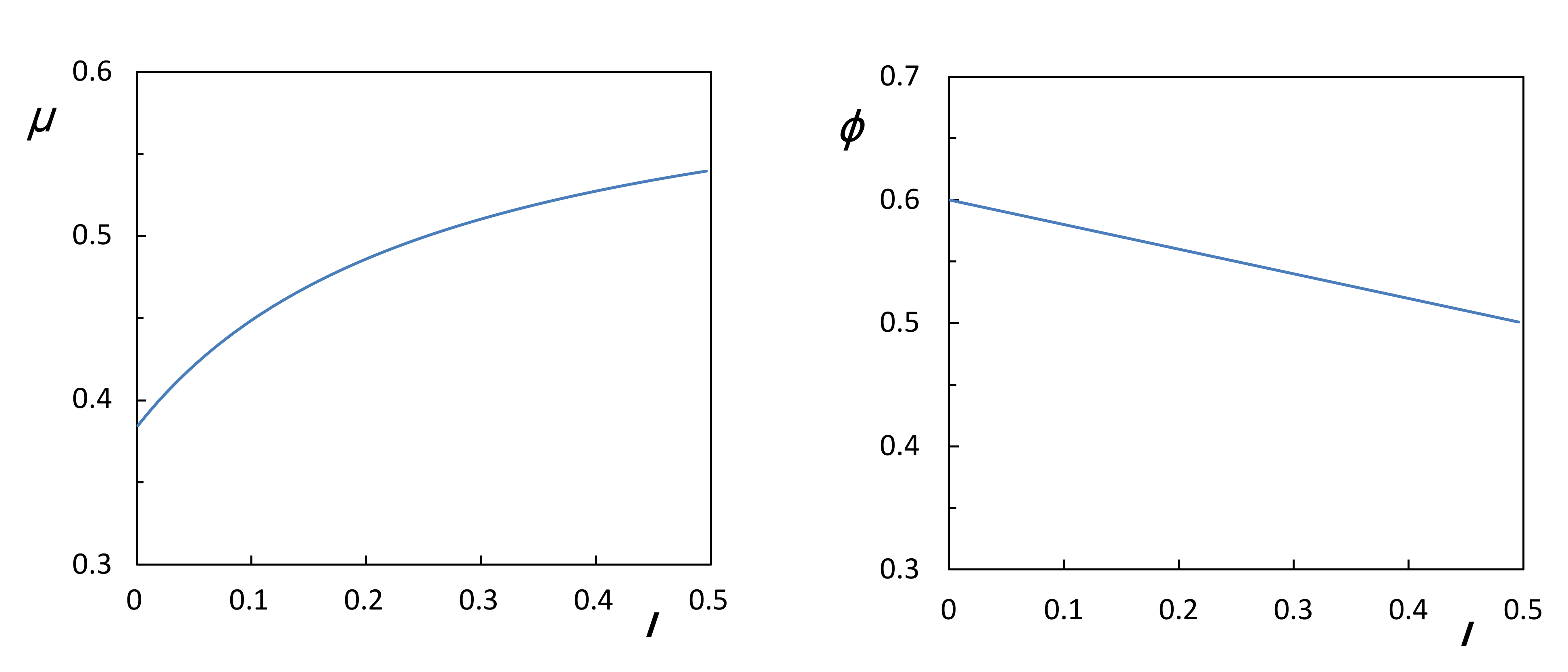}
	\caption{Schematic of the evolution of the friction coefficient $\mu$ (left) and of the volume fraction $\phi$ (right) as a function of the inertial number $I$.}
	\label{fig:muI}
\end{figure}

Assuming a collection of rigid frictional spheres in a plane shear geometry, one can neglect the elasticity. Then using dimensional arguments, one can define only two dimensionless quantities: the ratio of shear stress and normal stress defining a friction coefficient $\mu$ and an inertial parameter $I$ \cite{MiDi2004,Forterre2008}, also known as Savage number \cite{Savage1984} or Coulomb number \cite{Ancey1999}:
\begin{equation}
	I = \frac{\dot\gamma d}{\sqrt{P/\rho_p}},
\label{eq:I}
\end{equation}
where $\dot\gamma$ is the shear rate, $d$ is the grain diameter, $P$ is the \textcolor{black}{constant} confining pressure and $\rho_p$  the grain density.
This number can be viewed as the ratio of two time scales: The first one associated to the macroscopic shear rate is the time needed  for a grain to overtake its neighbor ($1/\dot\gamma$). The second is associated to the microscopic rearrangement, it is the time required for a grain to fall behind its neighbor under the confining pressure ($\sqrt{d^2\rho_p/P}$).
Decreasing $I$, the system gets closer to the quasi-static regime, whereas the kinetic regime is reached for high value of $I$.
Having only two dimensionless parameters, the friction coefficient $\mu(I)$ should be only a function of $I$. This was indeed proved by discrete numerical simulations of plane shear \cite{Dacruz2005,Iordanoff2004}.
For the dense flow regime,  corresponding to moderate inertial number (from 0.01 to 0.5), the friction coefficient $\mu(I)$ and the volume fraction $\phi(I)$ can be modeled accurately by the following phenomenological equations:

\begin{equation}
\mu = \mu_s + \frac{\mu_g-\mu_s}{I_0/I+1} \quad \textrm{and} \quad \phi = \phi_{\textrm{\small{max}}}-\Delta_\phi I,
\label{eq:muIphiI}
\end{equation}
where the constants are material dependent, $I_0\approx0.3$, $\Delta_\phi\approx0.2$. Figure (\ref{fig:muI},left) shows the typical behavior of the friction coefficient $\mu$ and volume fraction $\phi$ with the inertial number $I$. The friction coefficient starts then from a minimum value $\mu_s$ which coincides with the internal Coulomb friction of the material (typically $\mu_s\approx\tan21^\circ$ for glass spheres), and saturates toward a value $\mu_g$ ($\mu_g\approx\tan32^\circ$).
 One can observed that experimentally, steady homogeneous flows with higher values than $I=0.5$ are difficult to achieved due to instabilities or transition to the kinetic regime (see section \ref{sec:discussion}) \cite{Forterre2003,Forterre2008}, \textcolor{black}{thus the validity of Eq.~\ref{eq:muIphiI} at high $I$ is questionable}. The asymptotic limit $\mu_g$ is \textcolor{black}{nevertheless} required to account for the development of a gaseous phase in front of an avalanche \cite{Pouliquen1999b} or on top of a flow over a steep slope ($\tan\theta >\mu_g$) \cite{Jop2006,Taberlet2003}.
For practical applications, since the volume fraction remains close to its maximum value, one can often neglect its variations as a first order approximation and use only the evolution of $\mu$ with $I$.

It is worth noting that these laws were discovered in the plane shear simulations when \textcolor{black}{the volume fraction was no longer set}. Instead, the pressure is controlled and the volume fraction is free to adapt in the system with the evolution of other parameters. On the contrary if the volume fraction is fixed and the normal pressure measured, this will fixed the value of $I$. Then Equation (\ref{eq:I}) shows that the normal pressure should scale with the square of the shear rate. Indeed these scaling laws were measured experimentally in a Couette geometry with a fixed volume \cite{Savage1984}. 

\subsection{"Local" rheology}

Assuming the previous equations govern the flow properties locally, one can predict the velocity profiles in other configurations. Let us first consider the inclined plane where the tangential to normal stress ratio is fixed by the slope. Using equations (\ref{eq:I},\ref{eq:muIphiI}) to predict the velocity profile, the so-called Bagnold's profile is recovered in quantitative agreement with measurements \cite{MiDi2004,Dacruz2005}. The second validation was the two-dimensional flow on a heap. The concave velocity profiles are recovered by this rheology if one takes into account the influence of the sidewall correctly \cite{Jop2005} (see section \ref{sec:surface}).

To describe 3D flows, such as flows on complex topography or with transverse variations, a tensorial form of eq. (\ref{eq:muIphiI}) was proposed  \cite{Jop2006}. The deviatoric part of the stress tensor $\sigma_{ij}=-P\delta_{ij}+\tau_{ij}$:

\begin{equation}
\tau_{ij}=\frac{\mu(I) P}{|\dot{\gamma}|}\dot{\gamma}_{ij} \quad \textrm{where} \quad |\dot{\gamma}|=\sqrt{\dot{\gamma}_{ij}\dot{\gamma}_{ij}/2}.
\label{eq:mutensor}
\end{equation}

To write the above equation, the co-linearity of the shear stress and strain rate tensors is assumed as suggested \textcolor{black}{in first approximation} by theoretical considerations and numerical simulations \cite{Goddard1986,Depken2007,Dijksman2010}. \textcolor{black}{A closer analysis of the numerical simulations reveals however that there is a non-zero second normal difference of roughly 10\% of the mean pressure (see \cite{Depken2007} and below). Let us note further that equations \ref{eq:muIphiI} and \ref{eq:mutensor} need a closure equation. If the pressure is not constant in time, those equations cannot be used to compute it. A simple assumption commonly made in numerical works is to consider the granular flow incompressible. On the other hand to capture specific volume-fraction effects, an additional equation is required for its evolution (e.g. for immersed dense flows \cite{Pailha2008}).}

The first experimental test of this 3D law was done with flows on heap between rough walls which produce strong transverse shears, concluding that the model captures accurately the transverse free-surface velocity profile \cite{Jop2006}. However, the model still predicts a finite thickness instead of the slow, exponentially decreasing flow deep in the pile.
Despite this limitation, these correct predictions allow to gather the bulk flow properties from different configurations in a same theoretical framework.
 
\subsection{Complex geometries}

The $\mu(I)$ rheology was derived from simple flows, and first validated using steady unidirectional flows.
In front of its successes, researchers have studied more complex dense granular flows: 
To test further the prediction of this rheology, two main approaches were used. A first one is to simulate the granular flow as a continuum media using Navier-Stockes equations \cite{Chauchat2010,Lagree2011}. However using this pressure-dependence rheology is difficult, mainly because of the divergence of the viscosity close to the arrest. A second approach involves using discrete element method (DEM) to access locally to the rheological parameters in the complex flow and look for relation between shear stress and strain rate. I present some recent works supporting the $\mu(I)$ rheology carried on less constrained flows.

The first question that arises is whether this law is still valid for time-dependent flows, i.e. for accelerating flows. Studying the starting of unidirectional flows on heap, it was shown that both the time evolution and the amplitude can be predicted by this model \cite{Jop2007}.
A more drastic configuration is the collapse of a granular column on a rough plane. This peculiar configuration presents very fast flows that never reach steady states with a phase of strong acceleration at the base of the column \cite{Balmforth2005,Lube2004,Staron2005}. Lacaze \textit{et al.} performed numerical collapses of column by DEM, then analyzed the stress and strain rate tensors. Plotting the local friction coefficients as a function of the local inertial number, they obtain a collapse of all their data \cite{Lacaze2009}, although they observed a misalignment between tensors of few degrees in the foot (the static part) of the column. Recently, using a continuum code, dynamics of the collapse of a 2D column was achieved. \textcolor{black}{The very good agreement of shapes and dynamics between the continuum simulations and DEM simulations confirms the validity of a frictional rheology. However the relevance of the shape of $\mu(I)$ was questioned since a model with a constant friction coefficient could produce fairly good results in this case \cite{Lagree2011}.}

Another free surface flow exhibiting spatial inhomogeneity is the rotating drum configuration. Cortet \textit{et al.} analyzed carefully the numerical discrete simulations and found contrasting results: for the fast thin flowing layer that accelerates and decelerates exchanging grains with the erodible bottom, the local rheology applies \cite{Cortet2009}. Deep in the drum however, the co-linearity assumption is no longer valid. The researchers attributed it to the evolving contacts anisotropy between grains during the rotation or to the transient time of arrest. Surprisingly, the invariant form of equation (\ref{eq:mutensor}), $|\tau|=\mu(I)P$, is observed to be valid down to very small values of $I$ ($10^{-5})$.

Two last works tested the $\mu(I)$ rheology in vertical chute. First, the silo discharge through a small aperture was simulated using an incompressible continuum code. Two major features of the silo drainage were recovered: the independence on the height of the flow rate and the so-called Beverloo scaling linking the flow rate to the outlet diameter \cite{Staron2014}. In a second simulation, the drag force acting on a horizontal rod across the chute was simulated with a continuum code \cite{Chauchat2014} showing that the scaling with the Froude number is in qualitative agreement with experiments for slow flows.
Let us mention that the incompressibility used by the numerical solvers is not always a valid assumption, especially near the aperture or downstream the rod where the pressure drops to zero. Including the volume fraction variations remains a numerical challenge.
It is worth noting that the precise shape of the law $\mu(I)$ was not thoroughly tested in these continuum codes, but only the frictional viscosity (and yield stress) and the saturation of the friction coefficient for higher $I$. \textcolor{black}{Finally, an important remark concerns the nature of the flows mentioned above: The strain rate remains moderate in all configurations $I<0.5$ (except in some parts of the column collapse), this issue will be discussed in the section \ref{sec:fastflow}.}

%
\subsection{Beyond the average flow}
Up to now, I focused only on the average properties of the granular flows. However, the grain trajectories fluctuate around the mean path. The rheology is unable to take them into account, but measurements shows evidence of evolution of their amplitude with respect to the shear rate \cite{Bocquet2001} or to the inertial number $I$ \cite{Dacruz2005}. The standard deviation of the velocity fluctuations may be a consequence of the local shear rate and simply be a passive scalar \cite{MiDi2004}. Nevertheless, understanding their origin will allow a correct description of the mixing by diffusion of the grains during the flow \cite{Lu2008}.
  
Moreover, correlation lengths on velocities \cite{MiDi2004,Pouliquen2004} or forces \cite{Lois2007,Gardel2009} were measured and were shown to increase sharply when approaching the slow dense regime.
Having correlations over longer and longer distances signs the beginning of observed deviations from the prediction of the rheology. These growing length scales are the base of new theoretical developments to better understand the granular properties and to extend the existing model toward slow flows. Some of them will be presented in the next section.



\section{"Non-local" rheology}
\label{sec:nonlocal}
\textcolor{black}{Let us first remind that the term \textit{non-local} rheology is used in this article to enhance the spatial correlations that govern the flow behavior, even if all equations must still be evaluated locally.}

\subsection{Evidence of coupling across space}
Most of the granular flows in geophysics or industry are rather thin compared to their lateral extend. It is therefore important to compare the prediction of the rheology to the observation.

The first signs of the failure of the rheology $\mu(I)$ can be seen in the velocity profiles in several configurations. The scaling law on inclined plane was shown to deviate from the Bagnold's scaling as the thickness of the flow decreases. The curvature of the velocity profile decreases and can be inverted \cite{Silbert2003,Rajchenbach2003,MiDi2004}. The granular flow exhibits then a transition from the Pouliquen's flows rules to a non-local behavior \cite{Deboeuf2006,Takagi2011}.

These behaviors become more important close to the jamming transition. Therefore, such effect will be more pronounced at the interface between a fast flow and a quasi-static zone. Indeed, free surface flows on erodible material (heap or drum) exhibit a creeping region where the velocity profile decays exponentially \cite{Komatsu2001,MiDi2004,Crassous2008}. This feature is neither captured by the model which instead predicts a finite depth \cite{Jop2006}.
The same results are observed in cylindrical Couette devices where the local friction coefficient is no longer a single function of $I$ but depends on the internal diameter in the quasi-static region \cite{Koval2009}. 
\textcolor{black}{Interesting new ideas came from split-bottom geometries} \cite{fenistein2003,Moosavi2013}, in which the stress map produces wide shear bands in slow flows that do not fall into the $\mu(I)$ rheology framework \cite{Jop2008,Dijksman2010}. Derived experiments and simulations have evidenced a mechanical coupling across space away from a shear band (e.g. following the movement of an intruder far from a shear band) \cite{Reddy2009,Nichol2010,Unger2010,Bouzid2013}.
It is worth noting that these previous behaviors become visible in the presence of large stress heterogeneities.


\subsection{Modeling flows with high stress heterogeneity}
Correlations have been incorporated in several models to reproduce the non-local effects. One idea is to assuming the dissipation involves more grains (arches or columns). For example, assuming the kinetic energy is dissipated fully non-locally (over the whole thickness), a linear velocity profile is predicted on inclined plane \cite{Rajchenbach2003},
Mills \textit{et al.} integrates viscous stress along force chains \cite{Mills2000}, Jenkins \textit{et al.} dissipates kinetic energy of column of grains \cite{Chevoir2001}.
Andreotti extends a single grain model, which already captures the essence of the $\mu(I)$ rheology, assuming also that collisions occur between columns of grains in the bulk \cite{Andreotti2007}. These situations lead to an increase of momentum exchange and finally modify the velocity profile. \textcolor{black}{While these models provide new features, they rest on the existence of the correlations and yet this experimental fact is only supported by few numerical evidences \cite{Baran2006} and is still matter of debate \cite{Reddy2010}.}

In the following paragraphs, I will focus on situations exhibiting both quasi-static and flowing regions.
When the stress inside the system is highly heterogeneous (silo, cylindrical Couette, shear plane under gravity), shear bands appear and confine the flow. Nevertheless, far from the shear band, the material is still agitated by mechanical fluctuations \cite{Reddy2009,Nichol2010,Unger2010,Bouzid2013}.

The main advantage of the approaches discussed below is the ability to describe both the quasi-static regime and the flowing one, while incorporating the \textcolor{black}{$\mu(I)$} rheology.
One idea is to assume that the local shear rate provides mechanical fluctuations that promote the movement of other grains \cite{Pouliquen2009}. The fluctuation intensity decreases exponentially with a single characteristic length scale. This model succeeds in predicting qualitative behaviors such as shear banding in silo or the change of concavity of the velocity profile on inclined plane.

Different recent approaches are inspired by works on soft matter \cite{Bocquet2009}, where elasto-plastic events allows the grains to flow below the yield stress, and they have been adapted to the rigid nature of grains. At the lower order, these models reduce to a second-order spatial derivative equation of an order parameter (e.g. fluidity or inertial number) with a diverging length scale approaching the critical yield stress \cite{Kamrin2012,Bouzid2013,Henann2013}. These models based on numerical evidence can predict accurately the velocity profiles in quasi-static regions and show indeed a diverging length scale.
It is worth noting that the scaling of this length $l$ is close to the one derived by analogy to the Prandlt mixing length \cite{MiDi2004}: Assuming a viscosity scaling with $\rho l(I)^2\dot\gamma^2$, equation (\ref{eq:I}) leads to $l(I)\approx\sqrt{\mu(I)}/I$.

The remaining question is now to find the physical origin of such correlation lengths. Several candidates are  identified arches, contact and force network anisotropy, swirling clusters \cite{Miller2013}, the coordination number \cite{Sun2011}. One strategy of ongoing researches to test further the rheology is to study the response to small perturbations of an agitated or sheared steady system \cite{Wandersman2014,Wortel2014}.
More details on the rheological properties of slow flows will be found in \cite{HeckeCR,RadjaiCR}.


\subsection{Flow threshold}
Another signature of the correlation lengths is the existence of flow thresholds \cite{MiDi2004}, such as the minimum height of a flow at a given slope on inclined plane $h_{\textrm{\scriptsize{stop}}}$, the minimum flow rate on a pile or the minimum rotating rate in a drum experiment. All these conditions are not captured by the \textit{local} rheology which predicts steady flows at thicknesses as low as possible.

To address the issue of the arrest of flows, it seems natural to introduce a length scale able to correlate the movements of the grains over the whole flow thickness. In the "self-activated flow" model \cite{Pouliquen2009}, the flow ceases when the amount of the fluctuation is not enough to sustain it due to the reduce thickness. However the hysteretic nature of the arrest and start of the flow is not captured. 
 The starting-angle dependence on the thickness of an avalanche has been recently addressed through the analysis of stable granular configurations with some success \cite{Wyart2009}.

Finally, these thresholds depend not only on the flowing material but also on the roughness of the base \cite{Pouliquen1999,Goujon2007,Weinhart2012b}. \textcolor{black}{Tuning the amount of injected energy at the base in DEM simulations can modify the flow threshold and the bottom shear rate, but not the upper part of the velocity profile \cite{Maheshwari2012granmat}. This dependence on the nature of the dissipation at the base is difficult to understand if the correlation-length is only governed by the bulk behavior. However this length may be defined locally and may depend on the local agitation as discussed in the previous section. This issue raises the difficulty of setting correctly the boundary conditions.} In the next section, the different kinds of interactions between grains and boundaries are presented.

\section{Surface rheology}
\label{sec:surface}

Except from some numerical simulations, in every experiments or processes, one has to deal with lateral walls, fixed bottom. So besides the bulk rheology, one has to understand the properties induced by these different boundaries. The behavior of granular matter contrasts with the one of classical fluid where a zero velocity condition is often achieved. Since there is no scale separation between the microscopic length scale (the grain) and the typical length of the flow, a change in the roughness of a boundary may dramatically modify the flow properties such as the slip velocity \cite{Zheng1998}.
Which property is governed by a given boundary? Is it the shear rate, the average slip velocity or the shear stress?

I review in the following the current knowledge on the influence of boundaries.

\subsection{Lateral confinement}

One of the most underestimated effects in the past years was the effect of the lateral boundaries. Although it was noticed early that the presence of smooth sidewalls modifies slightly the flow regime \cite{Roberts1969
}, understanding their effect was the key feature to apply correctly the $\mu(I)$ rheology to the heap configuration \cite{Jop2005}. 
It was shown at the first order that a smooth vertical wall adds a frictional force to the momentum balance. Taking it into account and assuming a minimum friction coefficient at the liquid/solid interface allows to predict first a finite thickness of the confined flow over a static pile and second the increase of the inclination when increasing the flow rate as observed experimentally and numerically \cite{Taberlet2003,Jop2005,Taberlet2006}. However the effective friction coefficient that one has to use is somehow lower than the microscopic friction coefficient for these materials \cite{Jop2005,Bi2005}. This means either that the friction is not fully mobilized, for example due to the rotation of particles, either that friction forces are not completely aligned with the flow direction. A convincing work by Richard \textit{et al.} showed that below the fast flowing layer on a heap, the effective friction coefficient at the wall decreases due to the random motion of the grains in the quasi-solid phase \cite{Richard2008}. Moreover they showed that the decay of the friction coefficient is linked to the evolution of the density in the very fast flowing layer and that the creeping flow corresponds to a glassy state. Thus the effective friction is not an intrinsic property of the grains but depends on the microscopic agitation. It is worth noting that the resulting characteristic length of the exponential decrease of the velocity also depends on the fast layer properties. It is then not obvious if the non-local models described in the previous section can capture this exponential tail.

On the other hand, the use of rough lateral wall to impose a zero-slip velocity seems easier. Jop \textit{et al.} shows that, although this condition is appropriate at the first order to describe confined heap flows, careful analysis of the experimental velocity profiles reveal a slight slip at the rough wall \cite{Jop2006}.

The main shear often comes from the bottom and  the sidewalls act as a perturbation of the main flow. This may explain why simple laws can be used in first approximation here. These puzzling effective friction coefficients or slip velocities have to be understood to address dense granular processes accurately.
Although boundary conditions can be derived for kinetic theories, I will now focus on the stresses and slip velocities from some experimental and numerical works for basal boundaries.

\subsection{Rough bottom surfaces}

First evidences of the influence of the basal roughness on the flow properties are founded in the inclined plane configuration. Changing both the size of the glued spheres on the plane and the one of the flowing grains modifies the $h_{\textrm{\scriptsize{stop}}}(\theta)$ curves \cite{Pouliquen1999,Dacruz2004PhD,Goujon2007}. The influence of the roughness on the threshold can be interpreted by the change of dilatancy \cite{Mills2000}.

Pouliquen first shown that his scaling law still apply: The average velocity scales with the inverse of $h_{\textrm{\scriptsize{stop}}}$ corresponding to a couple grain-bottom. However, simulations of inclined plane have shown that the bulk flow properties are not influenced by the boundary conditions \cite{Mitarai2005}.
Moreover contrasting results exist in the literature: Shear plane simulations exhibit a rather homogeneous inertial number while inclined plane simulations show a strong variation of the inertial number close to the boundary \cite{Dacruz2003,Weinhart2012b}.
It is thus not clear if changing the roughness of the bottom only modifies the velocity and shear in a boundary or if it changes the whole bulk rheology \cite{Goujon2007} and so if the link with $h_{\textrm{\scriptsize{stop}}}$ is a coincidence or not. Non-local rheologies would provide some answers.

The role of the roughness has been thoroughly studied measuring the movement of a single grain down on inclined plane. An analytical 2D model captures the same phenomenology than the rheology $\mu(I)$ \cite{Andreotti2007}. If disorder is introduced in the position of the glued spheres but keeping the same density, then the same mean velocity is recovered \cite{Dippel1996,Dippel1997}.

Koval \textit{et al.} \cite{Koval2011,Koval2008PhD} shows a sharp increase of the apparent friction coefficient at the wall when increasing the roughness above 0.4 then a saturation. Moreover, the slip velocity increases when decreasing the roughness, in agreement with previous results \cite{Dacruz2004PhD}. 

Beside the slip velocity, changing the roughness also modifies the structure of the flow. When the ratio of the grain diameter at the boundary and in the flow decreases below 0.61, a transition occurs between a disordered system to an organized state in compact layers \cite{Kumaran2012,Kumaran2013}. Moreover, if the rough bottom is made of a regular dense ordered packing, oscillations between layered and disordered systems can be observed \cite{Silbert2002}.

Finally, the microscopic friction coefficient are shown to play a role only for small roughnesses \cite{Koval2008PhD,Shojaaee2012a}, while the increase of the rolling friction coefficient can trigger a transition between a sliding mode to a rolling one \cite{Estrada2008}.

\subsection{Smooth bottom surfaces}
It is expected that changing a rough bottom plane to a smooth one induces a slip velocity at the interface. However, the bulk properties such as the shear rate and volume fraction do not seems to be affected in simulations \cite{Artoni2012,Delannay2007}. Let us then focus on the interface properties.

For very low values of wall-particle friction coefficient $\mu_{wp}$, the material slips as a solid block. Increasing $\mu_{wp}$ above a critical value leads to the development of a shear layer \cite{Shojaaee2012a}. A first surprising result is that this critical value is lower than the particle-particle friction coefficient. The second results is that the effective friction coefficient (ratio of tangential stress to normal stress) drops significantly mimicing the static/dynamic transition for classical solid friction \cite{Shojaaee2012b}.
In a smooth shear plane configuration, they identified a boundary inertial number which governs the friction coefficient and three different regimes are observed: a linear velocity profile above an inertial number of 0.01, two symmetrical sheared layers at intermediate velocities and a single shear band regime for very low velocities.



Even though the link with the bulk inertial number is still lacking, an interesting step has been made to relate the slip velocity to the physical properties. Artoni \textit{et al.} studied the movement of a grain submitted to a stochastic force from an imaginary bulk and inferred a scaling law for the slip velocity \cite{Artoni2009PRE}. 
This empirical scaling has been confirmed analyzing flows of angular grains down frictional smooth plane numerically \cite{Artoni2012}. The slip velocity is governed by the ratio of the effective wall friction over the wall-particle friction coefficient $\mu_{eff}/\mu_{wp}$. It is worth noting that their law can be interpreted as a hydrodynamic slip length that diverges when going closer to 1 in the similar way that the diverging correlation lengths presented in the non-local continuum models.



Let us finally notice that a smooth frictional wall cannot model a rough plane: \textcolor{black}{The two effective friction coefficients coincide only in the limiting cases of very large or very small roughness \cite{Shojaaee2012b}. This result reminds of the complex role of the roughness already mentioned in the flow threshold section.}



\subsection{Erodible interface}

In the case of a flow over an erodible bottom, different approaches are possible. As shown previously \cite{Jop2006,Jop2007}, if the ground is composed of the same material then the bulk rheology could be used to find a first estimate of the evolution of the interface with time. Two difficulties arise. Since these flows are unsteady, there is no simple law linking the flow properties and the position of the interface and one has to compute it numerically. Moreover this "interface" separates fast and quasi-static flows, then the non-local effects of slow flows are more pronounced (see section \ref{sec:nonlocal}). 

In the following I briefly discuss an alternative approach to overcome this issue integrating the momentum and mass equations over the height.
When considering avalanches, an additional equation is required to couple the interface and flow properties evolutions. A first approach is to set the amplitude of exchange rate with the distance to a neutral inclination \cite{Bouchaud1994,Boutreux1998,Aradian2002}. Another idea is to impose a property of the flow (the velocity profile or the basal shear stress) then moving the interface modifies both the height of the flow and its average velocity \cite{Douady1999,Khakhar2001,Douady2001}.
Such method is useful for example to describe the wave instability observed at high flow rate \cite{Forterre2006} or to study the mobilization of a resting granular bed during the transient regime of an avalanche \cite{Mangeney2010}.  However the quantitative link with the bulk rheology is still lacking and required more effort on deriving analytical models.
The case of avalanches on very shallow flows \cite{Borzsonyi2005,Malloggi2006} can be described using a solid/liquid mixture model whose ratio is governed by an a Landau equation on the volume fraction \cite{Aranson2006}. It could be very fruitful to try to model them using the recent theoretical developments.


A last case of importance for the geophysical applications is when the ground has a given cohesion. Not only the average rheology at the interface is crucial to predict the velocity profile, the ground behaving as a rough solid bottom, but also the fluctuations of forces are important to understand the rate of the erosion process \cite{Lefebvre2013}. Understanding the physical background of these fluctuations, the impacts of grains, will also provide answers on the origin of the rheological law \cite{Yohannes2012,McCoy2013}.

                                          
\section{Discussion and conclusions}
\label{sec:discussion}

\subsection{Slow deformations}
So far we have seen that the rheology $\mu(I)$ presents a minimal relation to describe a large variety of dry granular flows. Moreover attempts have been made to extend the domain of application or to include more mechanical behaviors: e.g. to combine this rheology to the $h_{\textrm{\scriptsize{stop}}}$ \cite{Pouliquen2009}, to identify a relevant parameter for a non-local rheology able to explain creeping flows far from high sheared areas \cite{Kamrin2012,Bouzid2013,Henann2013}.

The remaining open question is the physical origin of these non-local behaviors. Obviously the microstructure is important.  The contact network and force anisotropies are shown to reflect the evolution of the friction coefficient even at high inertial number $I$ in shear plane flows \cite{Azema2014}. However, it is not clear whether such consideration could help explaining non-local effects: Which parameter governs the dynamics in this case?
 Minute variations of the volume fraction, assumed to be a passive parameter linked to $I$, could play a role in slow flows. The importance of the coordination number in the finite size effects on inclined plane ($h_{\textrm{\scriptsize{start}}}$) has been also pointed out \cite{Wyart2009}.

Although such approaches present a substantial improvement, some properties of the slow granular flows would still not be captured.  In the inertial regime only slight normal stress differences have been identified \cite{Silbert2001}. However, evidences of these effects can be found in slow flow experiments, such as the levee formation \cite{Felix2004,Deboeuf2006,Takagi2011} or free-surface transverse curvature \cite{McElwaine2012}. \textcolor{black}{These effects can be casted into the local 3D rheology \cite{McElwaine2012} and the depth-average predictions are in good agreement with the experimental surface profiles. Further tests of their model are now required on 3D flows}.

Finally, let us mention theoretical developments of more general mechanical models for plastic deformations in granular materials \cite{Goddard2014} that could serve as guide to identify the key parameters.



\subsection{Fast flows}
\label{sec:fastflow}
The other limit of the $\mu(I)$ rheology is the fast flow regime when part of the system becomes dilute. Transition toward salting grains are observed when the shear to normal stress ratio exceeds the limit $\mu_g$ \cite{Pouliquen1999,Taberlet2003,Jop2006}.
Numerical results show that beyond a critical value, an unstable branch of the friction coefficient exists: The friction coefficient starts decreasing with $I$ for greater values than 0.8, corresponding to the beginning of the dilute regime \cite{Lois2006PhD,Forterre2008,Borzsonyi2009}. The precise value of the transition depends on the softness of particles \cite{Chialvo2012}. These finding are in agreement with the low friction coefficient found in the single-grain model when the grain accelerates on a bumpy surface \cite{Andreotti2007} and with the intermediate friction coefficient for numerical supported flows \cite{Taberlet2007}. 
The description of very dilute flows relies on the physics of binary collisions and predicts indeed a decrease of the friction coefficient with the inertial number.
\textcolor{black}{However, dense stationary flows are still observed experimentally at high inclination, either in a confined configuration \cite{Taberlet2003} or for shallow flows on plane \cite{Holyoake2012}. These flows strongly challenge the rheology:
In between the dense regime at moderate inclinations and the complete dilute regime, the flow remains relatively dense and large deviations from the predictions of the $\mu(I)$ rheology are measured with higher friction coefficients \cite{Holyoake2012,Schaefer2013}.  
Aside from the stationary flows, instabilities often arise from the spatial variation of density and these new patterns, out of reach of the $\mu(I)$ rheology, strongly narrow the range of application of the rheology and can serve to test the kinetic models \cite{Forterre2002,Borzsonyi2009,Holyoake2012,Brodu2013}. One crucial issue would be to identify the nature of the dissipations of energy in such flows: e.g. inelastic collisions, secondary flows, "turbulent" flows.}


To adapt the kinetic theories to the dense regime, \textcolor{black}{researchers have relaxed the assumptions that the flow is dominated by binary collisions, since neglecting long lasting frictional contact may introduce large errors \cite{Lois2008}}.
Extensions to adapt the kinetic models to the dense regime provide finally a correct diverging thickness of $h_\textrm{\scriptsize{stop}}$ for low inclination \cite{Louge2003,Kumaran2008} and different features of inclined plane and heap flows can be well modeled \cite{Jenkins2006,Berzi2011}. However the flow is still described as the spatial juxtaposition of different regimes (base, dense core, diffuse free surface). 

More details on the recent developments of rheological models based on kinetic theories will be found in \cite{KumaranCR}.



\subsection{Complex interactions}
The main focus of this article was on the rheology of monodisperse dry spherical grains. What do these behaviors become if the interaction law between grains changes? As soon as we introduced a new physical ingredient, $I$ is no more the single dimensionless parameter in the system. Several works have tried to understand the rheology of polysdisperse systems, cohesive grains or non-spherical ones.

Pouring grains of different density or size on an inclined plane leads generally to a segregation phenomenon. Even though the mechanisms of the segregation are still matter of debate \cite{Daniels2013}, recent studies on binary dense flows show the validity of the local rheology in steady states. The friction coefficient is shown to follow the same dependence with the inertial number provided that the definition of $I$ is modified. One should replace the grains diameter by the average one \cite{Rognon2007,Yohannes2010} and the particle density by its mean value \cite{Tripathi2011}.

Also the nature of the interaction can be tuned. Attractive forces can arise when decreasing size or adding capillary bridges. The rheology of cohesive grains follows the same trend of increasing friction coefficient with $I$, however the cohesion increases globally its values \cite{Rognon2008,Brewster2005}. This effect is linked to the modification of the microstructure with an unexpected role of the distant attractive forces \cite{Khamseh2013a}.

When the grains are immersed in a viscous fluid, the same frictional behavior is observed and can be described provided to change the microscopic inertial time scale by a viscous one \cite{Pouliquen2006}. Aside the dynamical effect of volume fraction evolution, the very same theoretical framework can be applied then to dry grains and dense suspensions \cite{Lemaitre2009,Boyer2011,Trulsson2012}.

A final parameter is the shape of the particle. Irregular grains, like sand, deviate from the prediction of the simple rheology. Elongate grains are shown to orientate with the shear direction \cite{Borzsonyi2013}. Although influence on packing fraction and on slow shear regime of non-spherical particles has been investigated, the dense flow of such particles deserves more attention in the future. 



\subsection{Conclusion}
I restricted the scope of this article to dense flows. I presented their mechanical properties in the bulk as well as near boundaries together with the most recent theories.
The average features of the dense granular flows are now well understood. Several models have been built progressively
to explain them. \textcolor{black}{The rather useful model $\mu(I)$, presented in the GdR MiDi's article ten years ago, can predict some of the basic behaviors of the dense granular flow at moderate strain rate.} Since then the number of configurations of greater
degrees of complexity that take place within the same theoretical framework has continued to increase. \textcolor{black}{However, some common geophysical flows or some industrial applications, such as fast dense avalanches or high-shear mixing processes, cannot be modeled accurately by this rheology. This is part of its} current limits that I have shown and that still require other theories to be understood.

To conclude, let us mention other open questions. 
When deforming a granular media very slowly, for example in a biaxial test, the material does not flow homogeneously but present instead localization of the shear rate. Those shear bands qualitatively differ from the one developing close to a wall or in the split bottom geometry. Up to now, they are not described by the same model as dense flows. The question is then to try to link the behavior at high and low strain rates.
How can the most recent theoretical developments describe time-dependent flows where the anisotropy of contact network may be enhanced, such as oscillatory flows or the time-evolution of forces around intruder passing through his own wake \cite{Guillard2013}?  
Finally, what are the physical origins of the efficient models presented is a crucial question.  Underlying the non-local effects, I hope these theories will motivate interesting works to enlighten the relevant parameters.

\section*{Acknowledgment}

I would like to thank V. Kumaran and J. McElwaine for their interesting comments and for their critical reading of the manuscript.

Ref. 87 was corrected after publication.



  \bibliographystyle{elsarticle-num} 
  \bibliography{crphysbibjop}





\end{document}